\documentclass[twocolumn,showpacs,superscriptaddress,amsmath,amssymb,aps,pra]{revtex4-1}
\usepackage{graphicx}
\usepackage{dcolumn}
\usepackage{amsmath,bm}
\usepackage{epstopdf}
\usepackage[mathlines]{lineno}
\usepackage{color}
\usepackage[colorlinks=true,linkcolor=blue,citecolor=blue]{hyperref}

\begin{document}
\title{Method for identifying electromagnetically induced transparency in a tunable circuit quantum electrodynamics system}
\author{Qi-Chun Liu}
\thanks{These authors contributed equally to this work.}
\affiliation{Institute of Microelectronics, Department of Microelectronics and Nanoelectronics and
Tsinghua National Laboratory of Information Science and Technology, Tsinghua University, Beijing 100084, China}

\author{Tie-Fu Li}
\thanks{These authors contributed equally to this work.}
\affiliation{Institute of Microelectronics, Department of Microelectronics and Nanoelectronics and
Tsinghua National Laboratory of Information Science and Technology, Tsinghua University, Beijing 100084, China}
\affiliation{Quantum Physics and Quantum Information Division, Beijing Computational Science Research Center, Beijing 100193, China}
\affiliation{RIKEN Center for Emergent Matter Science (CEMS), 2-1 Hirosawa, Wako, Saitama 351-0198, Japan}

\author{Xiao-Qing Luo}
\affiliation{Quantum Physics and Quantum Information Division, Beijing Computational Science Research Center, Beijing 100193, China}

\author{Hu Zhao}
\affiliation{Institute of Microelectronics, Department of Microelectronics and Nanoelectronics and
Tsinghua National Laboratory of Information Science and Technology, Tsinghua University, Beijing 100084, China}

\author{Wei Xiong}
\affiliation{Quantum Physics and Quantum Information Division, Beijing Computational Science Research Center, Beijing 100193, China}
\affiliation{Department of Physics, Fudan University, Shanghai 200433, China}

\author{Ying-Shan Zhang}
\affiliation{Institute of Microelectronics, Department of Microelectronics and Nanoelectronics and
Tsinghua National Laboratory of Information Science and Technology, Tsinghua University, Beijing 100084, China}

\author{Zhen Chen}
\affiliation{Quantum Physics and Quantum Information Division, Beijing Computational Science Research Center, Beijing 100193, China}

\author{J. S. Liu}
\affiliation{Institute of Microelectronics, Department of Microelectronics and Nanoelectronics and
Tsinghua National Laboratory of Information Science and Technology, Tsinghua University, Beijing 100084, China}

\author{Wei Chen}
\thanks{weichen@tsinghua.edu.cn}
\affiliation{Institute of Microelectronics, Department of Microelectronics and Nanoelectronics and
Tsinghua National Laboratory of Information Science and Technology, Tsinghua University, Beijing 100084, China}

\author{Franco Nori}
\affiliation{RIKEN Center for Emergent Matter Science (CEMS), 2-1 Hirosawa, Wako, Saitama 351-0198, Japan}
\affiliation{Physics Department, The University of Michigan, Ann Arbor, MI 48109-1040, USA}

\author{J. S. Tsai}
\affiliation{RIKEN Center for Emergent Matter Science (CEMS), 2-1 Hirosawa, Wako, Saitama 351-0198, Japan}
\affiliation{Department of Physics, Tokyo University of Science, Kagurazaka, Shinjuku-ku, Tokyo 162-8601, Japan}

\author{J. Q. You}
\thanks{jqyou@csrc.ac.cn}
\affiliation{Quantum Physics and Quantum Information Division, Beijing Computational Science Research Center, Beijing 100193, China}

\begin{abstract}
Electromagnetically induced transparency (EIT) has been realized in atomic systems, but fulfilling the EIT conditions for artificial atoms made from superconducting circuits is a more difficult task. Here we report an experimental observation of the EIT in a tunable three-dimensional transmon by probing the cavity transmission. To fulfill the EIT conditions, we tune the transmon to adjust its damping rates by utilizing the effect of the cavity on the transmon states. From the experimental observations, we clearly identify the EIT and Autler-Townes splitting (ATS) regimes as well as the transition regime in between. Also, the experimental data demonstrate that the threshold $\Omega_{\rm AIC}$ determined by the Akaike information criterion can describe the EIT-ATS transition better than the threshold $\Omega_{\rm EIT}$ given by the EIT theory.
\end{abstract}
\date{\today}
\maketitle

\section {Introduction}

Driving a quantum three-level system with two resonant electromagnetic fields can induce destructive interference between different excitation pathways. Known as the electromagnetically induced transparency (EIT) (see, e.g., Refs.~\cite{Marangos1998,Fleischhauer2005}), this important effect can be used to slow down and even stop or trap optical~\cite{Hau1999,Kash1999} and microwave photons~\cite{Capmany2011,Zhou2013}.
Also, it has potential applications in single-photon storage~\cite{Liu2001,vanderWal2003,Kuzmich2003,Chaneliere2005} and can be used to achieve a quantum transistor by combining it with cavity quantum electrodynamics (QED)~\cite{Souza2013}. In fact, EIT has been realized experimentally in various systems, e.g., atomic~\cite{Lukin1997,Field1991} and molecular systems~\cite{Mucke2010,Tamarat1995}, quantum dots~\cite{Xu2007,Xu2008} and whispering-gallery-mode microresonators~\cite{Peng2014}. As one of the most promising systems for implementing quantum information processing, superconducting quantum circuits can also be used to demonstrate quantum-optics phenomena and effects occurring in atomic systems~\cite{You2011}.

Resulting from Fano interference~\cite{Fano1961} between two field-induced transitions, EIT creates a transparency window in the measured absorption or transmission spectrum of the system. A transparency window can also be created by Autler-Townes splitting (ATS). Instead of being due to the interference effect, it is caused by the electromagnetic-pumping doublet structure in the absorption or transmission spectrum~\cite{Autler1955}. Because of the similar transparency windows in the spectrum, EIT has often been confused with ATS. In the field of superconducting quantum circuits, there have been some experiments~\cite{Sung2014,Novikov2013,Li2011,Hoi2011,William2010,Abdumalikov2010,Mika2009} involving either ATS or EIT. Theoretical analyses indicated that the claimed EIT was actually ATS~\cite{Anisimov2011,Sun2014}. This is because of the difficulty for these superconducting circuits to satisfy the damping-rate conditions for realizing the EIT in experiments~\cite{Sun2014}.
Therefore, it remains an unsolved, important problem to realize EIT in a superconducting quantum circuit.

In this paper, we report an experimental observation of the EIT in a circuit quantum electrodynamics system consisting of a transmon qubit and a three-dimensional (3D) waveguide cavity. The key point is to engineer a tunable effective environment for the transmon states by utilizing the effect of the cavity, which is impossible when using an open system such as an open coplanar waveguide~\cite{Hoi2011,Abdumalikov2010}. By varying the magnetic flux in the superconducting quantum interference device (SQUID) loop to tune the transition frequency between the ground state and the second excited state of the transmon, we can adjust the damping rates between transmon states to reach the EIT regime of the system.
Indeed, both our experimental results and the Akaike information criterion (AIC)~\cite{Anisimov2011} analysis of the measured cavity transmission spectrum explicitly reveal that the 3D transmon system has reached the EIT regime in certain situations.

\section {Experiment}

\begin{figure}
\scalebox{.5}{\includegraphics{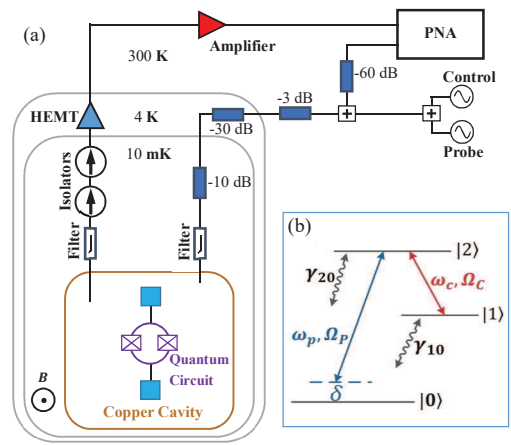}}
\caption{\label{fig1}(color online) (a) Schematic diagram of the experimental setup. A tunable 3D transmon consisting of a symmetric SQUID and a copper 3D cavity is thermally anchored to the mixing chamber, biased with a static magnetic field. A network analyzer works at 8.21950 GHz (the resonant frequency of the 3D cavity when the transmon is in the ground state $|0\rangle$) with a fixed power of -15 dBm at its output port, corresponding to the average photon number in the cavity to be $\sim$ 0.7. At this frequency, the transmission coefficient $T$ of the cavity is measured. A microwave source provides the control tone at $\omega_c=\omega_{21}=2\pi\times3.97950$ GHz with various powers. Another microwave source provides the probe tone at $\omega_p=\omega_{20}-\delta$ with a fixed power of -50 dBm at the source output. The microwave signals are combined at room temperature by two splitters and then strongly attenuated and filtered before reaching the sample. The transmitted signal through the 3D cavity is amplified at 4K and room temperature before received by the network analyzer. (b) The lowest three energy levels of the transmon driven by a control field (red) and a probe field (blue). The control field with frequency $\omega_c$ is in resonance with the transition between $|1\rangle$ and $|2\rangle$, and the probe field with frequency $\omega_p$ has a detuning $\delta$ with the transition between $|0\rangle$ and $|2\rangle$. The corresponding driving strengths are $\Omega_c$ and $\Omega_p$, respectively.}
\label{figure1}
\end{figure}

The device used is a tunable 3D transmon, 
where the single Josephson junction in a conventional 3D transmon~\cite{Paik2011, Rigetti2012} is replaced by a SQUID with two identical Josephson junctions. This symmetric SQUID is fabricated on a silicon substrate using the standard double-angle evaporation process; the Al/AlOx/Al junction has an area of $140$~nm$\times150$~nm, the SQUID loop is of the size $2~\mu$m$\times 4~\mu$m, and each shunting capacitor Al pad has an area of $250~\mu$m$\times500~\mu$m. These two Al pads and the cavity constitute a large capacitance shunted to the SQUID. The charging energy of this transmon is measured to be $E_C/h=412$~MHz, which includes the effect of the shunt capacitance. The SQUID behaves as an effective Josephson junction tuned by the externally applied magnetic field. At the bias magnetic field where the EIT occurs, the coupling energy of the effective Josephson junction is measured to be $E_J/h=7.0$~GHz. The 3D cavity has dimensions $40.0$~mm$\times21.0$~mm$\times4.5$~mm, with a fundamental eigenmode TE$_{101}$ of $\omega_{\textrm{cavity}}/2\pi=8.21690$~GHz and a loaded quality factor $Q_L\approx1000$. The coupling strength between the first excited state of the transmon and the cavity mode is $g/2\pi=173$~MHz, as obtained here via the vacuum Rabi splitting measurement.
The experiment was performed in a BlueFors LD-400 dilution refrigerator at $\sim 25$~mK [see Fig.~\ref{figure1}(a)].

We use $|i\rangle$, $i=0,1,2$, to denote the lowest three eigenstates of the transmon with the corresponding energies $\hbar\omega_i$ [see Fig.~\ref{figure1}(b)], where the transition frequency between states $|i\rangle$ and $|j\rangle$ is $\omega_{ij}=\omega_i-\omega_j$ ($i > j$).
The cavity acts as an effective environment for the transmon states and this effective environment can be engineered to be tunable by varying the detuning of the transition frequency $\omega_{20}$ from the cavity frequency via the magnetic field in the SQUID loop. Notably, our experimental results show that when decreasing this frequency detuning, the damping rate $\gamma_{20}$ between $|2\rangle$ and $|0\rangle$ can be greatly increased.
This is due to the enhanced dissipation from the cavity as the transition frequency $\omega_{20}$ becomes close to the cavity frequency. Here the cavity is used to engineer the noise spectrum of the transmon, where the noise around the cavity frequency can be stronger. Moreover, as can be seen from the Appendix, the dominating noise channel for $\gamma_{20}$ is due to the flux noise through the SQUID loop. Because the transmon is detuned from the flux sweet spot in the present case, its sensitivity to the flux noise is also increased.
To determine the damping rate $\gamma_{10}$ ($\gamma_{20}$) between $|1\rangle$ ($|2\rangle$) and $|0\rangle$, various spectroscopy tones with different strengths have been applied to measure the spectroscopy of the transmon. While the spectroscopy tone is weak enough (for this experiment, the spectroscopy tone strength is estimated to be $\Omega_{p}/2\pi=0.35$~MHz), the nearly intrinsic linewidth can be obtained.
By fitting each peak in the measured transmission spectrum of the cavity via a Lorentzian, we obtain the damping rates $\gamma_{10}/2\pi=1.76$~MHz and $\gamma_{20}/2\pi=6.90$~MHz.  
Also, the transition frequencies are measured to be $\omega_{10}/2\pi=4.39125$~GHz, $\omega_{20}/2\pi=8.37075$~GHz, and $\omega_{21}/2\pi=3.97950$~GHz. The detuning of $\omega_{20}/2\pi$ from the cavity frequency is 153.85~MHz.

In the experiment for EIT and ATS, we applied a control field in resonance with the transition frequency between $|2\rangle$ and $|1\rangle$ (i.e., $\omega_c=\omega_{21}$) and a probe field slightly detuned with the transition frequency between $|2\rangle$ and $|0\rangle$ (i.e., $\omega_p=\omega_{20}-\delta$) [see Fig.~\ref{figure1}(b)].
The corresponding driving strengths of these two microwave tones on the three-level system are $\Omega_c$ and $\Omega_p$, respectively. While these probe and control tones interact with the three-level system, they cannot transmit through the cavity due to the off-resonance with the cavity. Also, because the cavity mode is only dispersively coupled to the three-level system, the quantum dynamics of this three-level system becomes effectively decoupled from the cavity mode in this dispersive regime.

\section {A driven three-level system dispersively coupled to a 3D microwave cavity}

As depicted in Fig.~\ref{figure1}(b), under the rotating-wave approximation, the Hamiltonian of the system reads $H=H_0+H_{\rm int}$, with (we set $\hbar=1$)
\begin{eqnarray}
H_0&=&\omega_{a} a^\dag a+\nu_{10}|1\rangle\langle1|+\nu_{20}|2\rangle\langle2|,\\
H_{\rm int}&=&g_1(a^\dag |0\rangle\langle1|+a |1\rangle\langle0|)+g_2(a^\dag |1\rangle\langle2|+a |2\rangle\langle1|),\nonumber
\end{eqnarray}
where $a$ ($a^\dag$) is the annihilation (creation) operator of the cavity mode, $\omega_{a}$ is the dispersive frequency of the cavity when the transmon is in the ground state, $\nu_{10}$ ($\nu_{20}$) is the level difference between the state $|1\rangle$ ($|2\rangle$) and the state $|0\rangle$, and $g_1$ ($g_2$) is the coupling strength between the cavity mode and the transition $|1\rangle\leftrightarrow|0\rangle$ ($|2\rangle$). Here we ignore the weak coupling between the transition $|2\rangle\leftrightarrow|0\rangle$ and the cavity mode because we observed no vacuum Rabi splitting when $\nu_{20}$ is resonant with the cavity frequency, as in Ref.~\onlinecite{baur}. The driven Hamiltonian is
\begin{eqnarray}
H_d=&-&(\Omega_c|2\rangle\langle1|e^{-i\omega_c t}+\Omega_p|2\rangle\langle0|e^{-i\omega_p t}\nonumber\\
&+&\Omega_m a^\dag e^{-i\omega_m t}+{\rm H.c.}),
\end{eqnarray}
where $\Omega_m$ is the coupling strength between the cavity mode and the measurement field with frequency $\omega_m$.

The total Hamiltonian of the system can be written as $H_{\rm tot}=H_0+H_{\rm int}+H_d$. In the considered dispersive regime, i.e.,
$|g_1/\Delta_{10}|\ll1$, and~$|g_2/\Delta_{21}|\ll1$,
with detunings $\Delta_{10}\equiv\omega_{a}-\nu_{10}$ and $\Delta_{21}\equiv\omega_{a}-\nu_{21}$, a Fr\"{o}hlich-Nakajima transformation can be employed to convert the total Hamiltonian to
\begin{eqnarray}
\mathcal {H}_{\rm tot}&=& UH_{\rm tot}U^\dag\nonumber\\
&=&\omega_{a}a^\dag a+\omega_{10}|1\rangle\langle1|+\omega_{20}|2\rangle\langle2|+H_d\nonumber\\
& &-[g_1\chi_1 |0\rangle\langle0|-(g_1\chi_1 -g_2\chi_2)|1\rangle\langle1|\nonumber\\
& &-g_2\chi_2|2\rangle\langle2|]a^\dag a,
\label{Heff}
\end{eqnarray}
where $\omega_{10}=\nu_{10}+g_1\chi_1$, $\omega_{20}=\nu_{10}+\nu_{21}+g_2\chi_2$, and the transformation is
$U=\exp(-V)$, with
\begin{equation}
V=\chi_1(a^\dag |0\rangle\langle1|-a |1\rangle\langle0|)+\chi_2(a^\dag |1\rangle\langle2|-a |2\rangle\langle1|).
\end{equation}
Here $V$ satisfies $H_{\rm int}+[H_0,V]=0$, which gives rise to $\chi_1=-g_1/\Delta_{10}$ and $\chi_2=-g_2/\Delta_{21}$. In Eq.~(\ref{Heff}), terms up to the second order are kept due to the small coefficients $\chi_1$ and $\chi_2$. Also, the weak two-photon processes are ignored. Because $\Omega_p,~\Omega_m,~\Omega_c\ll g_1,~g_2$ in our experiment, the interaction Hamiltonian $H_d$ is approximately unaffected by the unitary transformation.

By further applying another unitary transformation,
\begin{equation}
S=\exp[-i(\omega_p-\omega_c)t|1\rangle \langle 1|-i\omega_p t |2\rangle \langle 2| -i\omega_m t a^\dag a],
\end{equation}
the Hamiltonian~(\ref{Heff}) is converted to
\begin{eqnarray}
\mathcal {H}&=&S^\dag \mathcal {H}_{\rm tot}S-iS^\dag\partial_t S\nonumber\\
&=& \delta_a a^\dag a+\delta_{10}|1\rangle\langle1|+\delta_{20}|2\rangle\langle2|-[g_1\chi_1 |0\rangle\langle0|\nonumber\\
&&-(g_1\chi_1 -g_2\chi_2)|1\rangle\langle1|-g_2\chi_2|2\rangle\langle2|]a^\dag a\nonumber\\
&&-(\Omega_c|2\rangle\langle1|+\Omega_p|2\rangle\langle0|+\Omega_m a^\dag +{\rm H.c.}),
\label{Heff'}
\end{eqnarray}
where $\delta_a=\omega_a-\omega_m$, $\delta_{10}=\omega_{10}+\omega_c-\omega_p$, and $\delta_{20}=\omega_{20}-\omega_p=\omega_{10}+\omega_{21}-\omega_p$. In our experiment, we have $\omega_c=\omega_{21}$, so $\delta_{10}=\delta_{20}\equiv\delta$.

The quantum dynamics of the system can be described by the following Born-Markov master equation:
\begin{align}
\frac{\partial\rho}{\partial t}=&-i[\mathcal{H},\rho]+\mathcal{L}[\rho]\nonumber\\
=&-i[\mathcal{H},\rho]+\frac{\Gamma_{10}}{2}\mathcal{D}[|0\rangle \langle1|]\rho+\frac{\Gamma_{20}}{2}\mathcal{D}[|0\rangle \langle2|]\rho\nonumber\\
&+\frac{\Gamma_{21}}{2}\mathcal{D}[|1\rangle \langle2|]\rho+\gamma_{00}^\phi\mathcal{D}[|0\rangle \langle0|]\rho+\gamma_{11}^\phi\mathcal{D}[|1\rangle \langle1|]\rho\nonumber\\
&+\gamma_{22}^\phi\mathcal{D}[|2\rangle \langle2|]+\frac{\kappa}{2}\mathcal{D}[a]\rho,
\label{master}
\end{align}
where $\mathcal {D}[\mathcal {O}]\rho=2\mathcal {O}\rho\mathcal{O}^\dag-\mathcal {O}^\dag\mathcal{O}\rho-\rho\mathcal {O}^\dag\mathcal{O}$, with $\mathcal{O}=a$ and $|l\rangle \langle m|$, $l\leq m\in\{0,1,2\}$. The corresponding coefficient of $\mathcal {D}[\mathcal {O}]\rho$ denotes the dissipation rate, including relaxation and pure dephasing rates. Note that this master equation applies to both the $\Lambda$- and $\Delta$-type three-level systems, which correspond to $\Gamma_{10}=0$ and $\Gamma_{10}\ne 0$ in Eq.~(\ref{master}), respectively.

From Eq.~(\ref{master}), we can explicitly write
\begin{align}
\frac{\partial \rho_{10}}{\partial t}&=-(\gamma_{10}+i\delta)\rho_{10}+i\Omega_c\rho_{20}-i\Omega_p\rho_{12},\nonumber\\
\frac{\partial \rho_{20}}{\partial t}&=-(\gamma_{20}+i\delta)\rho_{20}+i\Omega_c\rho_{10}-i\Omega_p(\rho_{22}-\rho_{00}),\label{eq1}
\end{align}
where $\gamma_{10}=\frac{1}{2}\Gamma_{10}+\gamma_{00}^\phi+\gamma_{11}^\phi$ and $\gamma_{20}=\frac{1}{2}(\Gamma_{20}+\Gamma_{21})+\gamma_{00}^\phi+\gamma_{22}^\phi$.
From Eq.~(\ref{eq1}), it can be seen that the quantum dynamics of the three-level system is effectively decoupled from the cavity mode in the dispersive regime that we considered.
For a weak probe field ($\Omega_p\ll\Omega_c$), which is valid in both EIT and ATS regimes in our experiment, starting from the ground state $|0\rangle$, the off-diagonal density-matrix element $\rho_{20}$ of the three-level system in the steady state can be obtained from Eq.~(\ref{eq1}) as
\begin{eqnarray}
\rho_{20}=\frac{\Omega_p}{\delta-i\gamma_{20}-\frac{\Omega_c^2}{\delta-i\gamma_{10}}}.
\label{rho20}
\end{eqnarray}
A similar result can also be found in Ref.~\onlinecite{Scully} for the $\Lambda$-type three-level system.

In a steady state, from Eq.~(\ref{master}), we can obtain
\begin{align}
\rho_{11}&=\frac{2C_1\Omega_p}{C_1\Gamma_{20}+C_2\Gamma_{10}}{\rm Im}(\rho_{20}),\nonumber\\
\rho_{22}&=\frac{2C_2\Omega_p}{C_1\Gamma_{20}+C_2\Gamma_{10}}{\rm Im}(\rho_{20}),
\end{align}
with the parameters $C_1=\gamma_{21}(\Gamma_{10}\gamma_{21}-2\Omega_c^2)$ and $C_2=\gamma_{21}(\Gamma_{21}\gamma_{21}-2\Omega_c^2)$.

\section {EIT in a tunable 3D cavity}

Here we apply a readout tone to the cavity at $8.21950$~GHz, which corresponds to the resonant frequency of the cavity when the three-level system is in the ground state $|0\rangle$. The measured transmission spectrum $T$ of the cavity depends on the occupation probabilities of the three-level system given by $T=\rho_{00}T_0+\rho_{11}T_1+\rho_{22}T_2$, where $\rho_{00}+\rho_{11}+\rho_{22}=1$, and $T_0$, $T_1$ and $T_2$ are the cavity transmission coefficients when the transmon is in the state $|0\rangle$, $|1\rangle$ and $|2\rangle$, respectively. Because $\rho_{11}\propto{\rm Im}(\rho_{20})$ and $\rho_{22}\propto{\rm Im}(\rho_{20})$, the normalized transmission coefficient is proportional to ${\rm Im}(\rho_{20})$:
\begin{eqnarray}
T'&\equiv&\frac{T-T_0}{T_2-T_0}=\frac{T_1-T_0}{T_2-T_0}\rho_{11}+\rho_{22}= A{\rm Im}(\rho_{20})\nonumber\\
&=&\frac{A\Omega_p\left(\gamma_{20}+\frac{\Omega_c^2\gamma_{10}}{\delta^2+\gamma_{10}^2}\right)}{\left(\delta-\frac{\Omega_c^2\delta}{\delta^2+\gamma_{10}^2}\right)^2+\left(\gamma_{20}+\frac{\Omega_c^2\gamma_{10}}{\delta^2+\gamma_{10}^2}\right)^2}.
\label{eq:t}
\end{eqnarray}

In quantum optics, the imaginary part of the complex susceptibility (${\rm Im}\chi\propto{\rm Im}(\rho_{20})$) is measured to demonstrate the EIT~\cite{Scully}, which is directly related to the probe tone applied to the atomic systems.
However, in the circuit-QED approach for superconducting circuits, the cavity transmission is usually measured.
Because $T'\propto{\rm Im}(\rho_{20})$ is obtained in our experiment, the measured normalized transmission coefficient $T'$ can also demonstrate the EIT in the 3D transmon.

\begin{figure}[!hbt]
\scalebox{.22}{\includegraphics{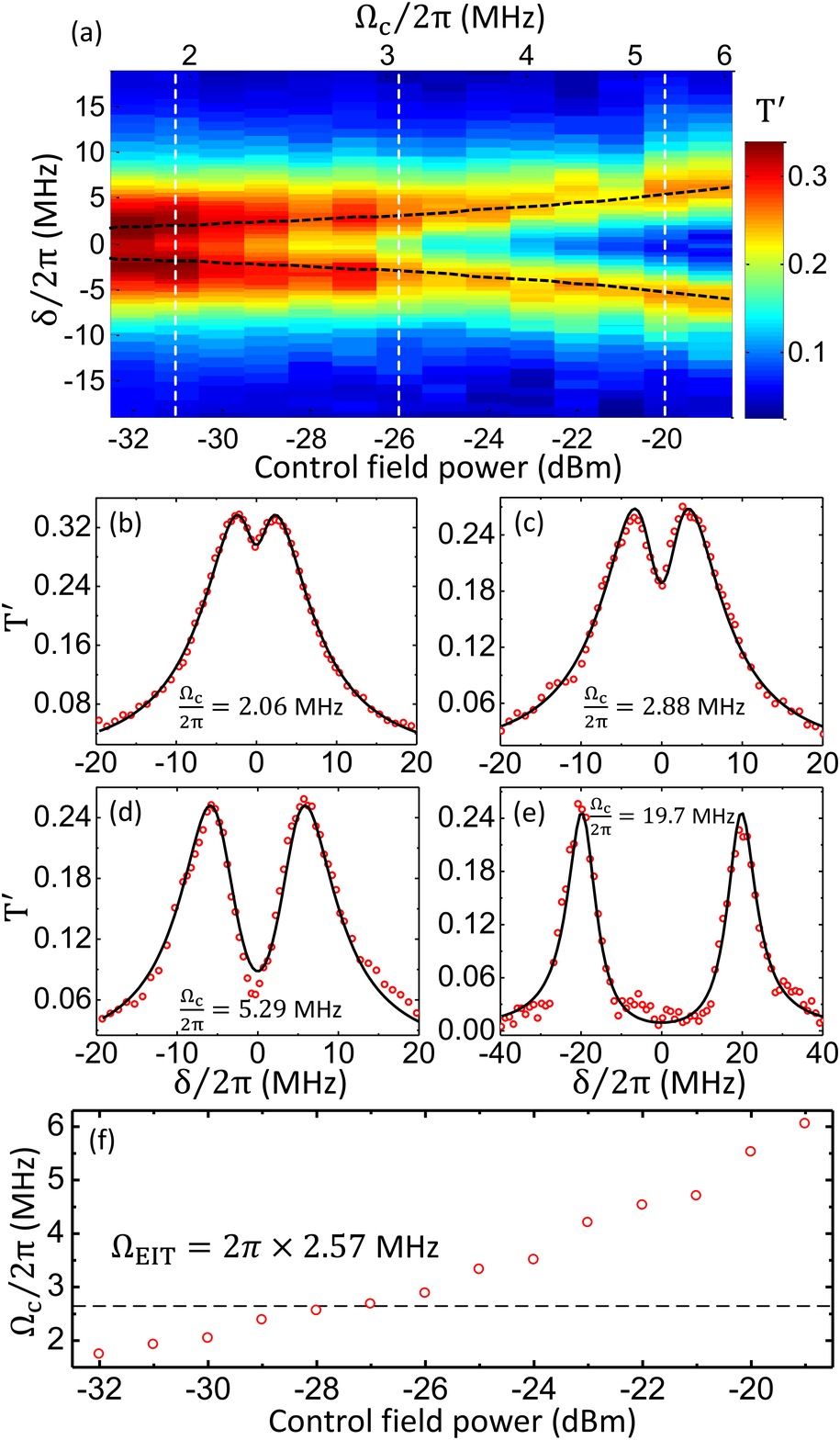}}
\caption{\label{fig2}(color online) (a)~The normalized transmission coefficient $T'$ of the cavity, where the control field sweeping from -32dBm to -19 dBm at source output is applied in resonance with the transition frequency $\omega_{21}/2\pi=3.97950$ GHz, and the probe field with power -50 dBm at source output has a detuning $\delta/2\pi$ from the transition frequency $\omega_{20}/2\pi=8.37075$ GHz. The dashed (black) guide lines correspond to the two peaks in $T'$. In order to clearly show the two peaks in the weak control-field range, we present $T'$ in (a) only up to -19dBm for the control-field power.
(b)-(e)~The measured transmission coefficient $T'$ versus the frequency detuning $\delta/2\pi$ (denoted by red open circles) at different powers of the control field: (b)~-31dBm, (c)~-26dBm, (d)~-20dBm and (e)~-8dBm, where the results at -31dBm, -26dBm and -20dBm correspond to the three vertical (white) lines in (a) and the result at -8dBm is not shown there. The solid curves are the fitting results obtained using Eq.~(\ref{eq:t}) when $\Omega_c/2\pi=2.06, 2.88, 5.29$ and $19.7$~MHz, respectively. (f)~The driving strength $\Omega_c$ at each control-field power (denoted by red open circles), which is obtained by fitting the measured transmission coefficient $T'$ with Eq.~(\ref{eq:t}). The upper limit of the driving strength $\Omega_{\rm EIT}$ for realizing EIT is indicated by the dashed line.}
\label{figure2}
\end{figure}

In Fig.~\ref{figure2}(a), we show $T'$ for different control-field powers, ranging from $-32$~dBm to $-19$~dBm at the source output. By fitting $T'$ in Fig.~\ref{figure2}(a) with Eq.~(\ref{eq:t}) at each control-field power, we obtain the driving strength $\Omega_c$ ranging from $2\pi\times1.74$~MHz to $2\pi\times6.05$~MHz. For example, Fig.~\ref{figure2}(b)-2(e) show the fitting of experimental data with Eq.~(\ref{eq:t}) at some typical control field powers: $-31$~dBm, $-26$~dBm, $-20$~dBm and $-8$~dBm, which correspond to $\Omega_c/2\pi=2.06$~MHz, $2.88$~MHz, $5.29$~MHz and $19.7$~MHz, respectively. For a three-level system driven as in Fig.~\ref{figure1}(b) by both a control field and a probe field, the conditions for realizing EIT (i.e., to create a dark state with only superposition of $|0\rangle$ and $|1\rangle$) are~\cite{Sun2014} $\gamma_{20}>2\gamma_{10}$, and
\begin{equation*}
\gamma_{10}\sqrt{\gamma_{10}/(2\gamma_{10}+\gamma_{20})}<\Omega_c<(\gamma_{20}-\gamma_{10})/2.
\end{equation*}

\noindent In our experiment, $\gamma_{20}/2\pi=6.90$~MHz and $\gamma_{10}/2\pi=1.76$~MHz (which satisfy $\gamma_{20}>2\gamma_{10}$), so the EIT conditions require that $0.79$~MHz $<\Omega_c/2\pi<2.57$~MHz. In Fig.~\ref{figure2}(f), it can be seen that part of the applied $\Omega_c$ is within the range that satisfies the EIT conditions.

\section {Discriminating EIT from ATS}

\begin{figure}
\scalebox{.22}{\includegraphics{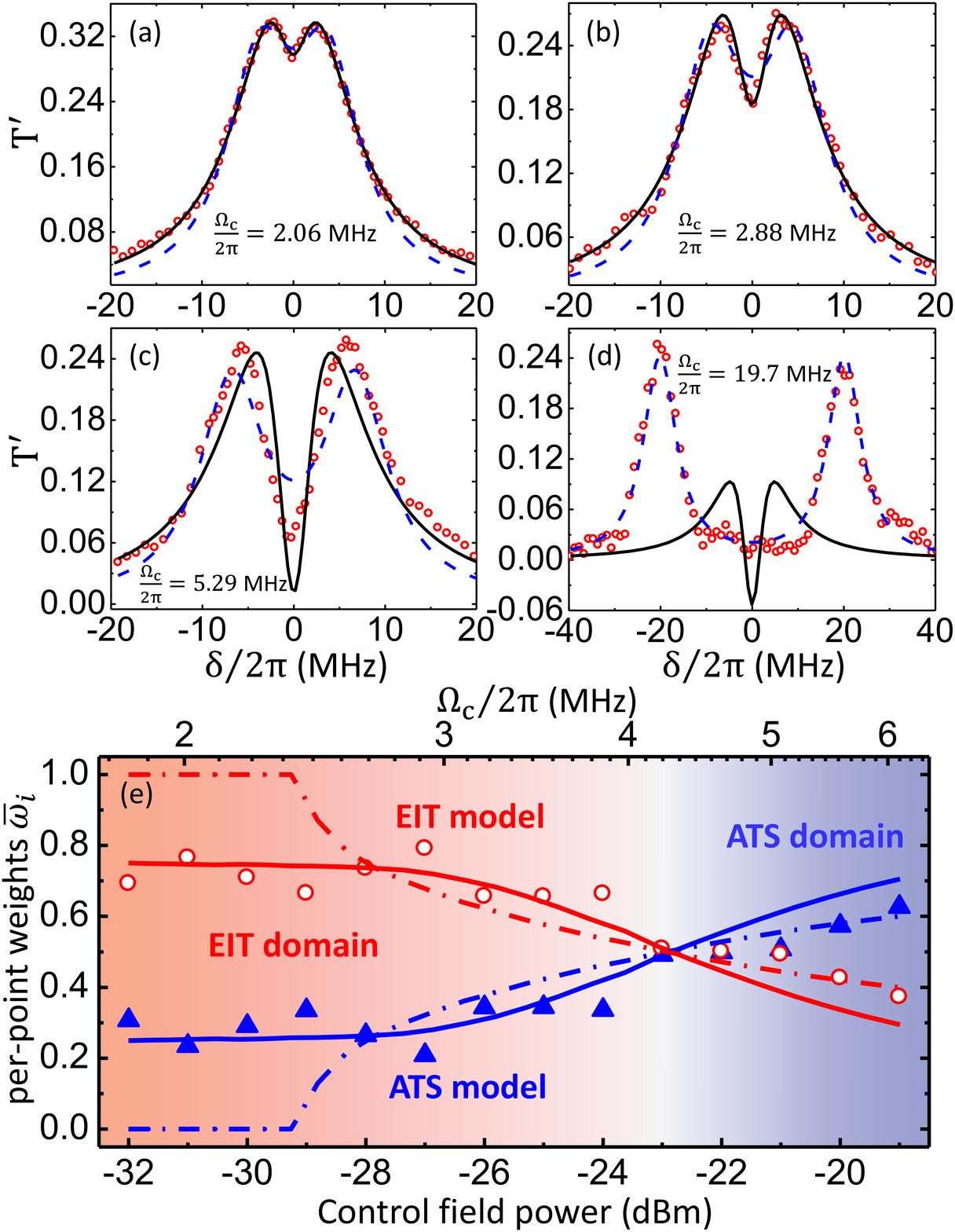}}
\caption{\label{fig3}(color online) (a)-(d)~Comparison of the measured transmission coefficient $T'$ (denoted by red open circles) with the results obtained using Eq.~(\ref{eq:eit}), i.e., the EIT model (solid curves) and Eq.~(\ref{eq:ats}), i.e., the ATS model (dashed curves).
The powers of the control field are (a)~-31dBm, (b)~-26dBm, (c)~-20dBm and (d)~-8dBm, which correspond to $\Omega_c/2\pi=2.06, 2.88, 5.29$ and $19.7$~MHz, respectively.
(e)~The AIC per-point weights for both EIT and ATS models at different values of the control-field power (i.e., different driving strength $\Omega_c$). The red open circles (blue solid triangles) correspond to the per-point weights of the EIT (ATS) model obtained by fitting with the experimental data. The red (blue) solid curve corresponds to the theoretical AIC per-point weights obtained with $\gamma_{20}/2\pi=6.90$ MHz and $\gamma_{10}/2\pi=1.76$ MHz as well as an additional $3\%$ experimental noise, while the red (blue) dash-dotted curve corresponds to the theoretical AIC per-point weights obtained without the additional experimental noise.}
\label{figure3}
\end{figure}

Below we further analyze our experimental results to see how the driven three-level system transitions from the EIT to ATS regime. When the system is in the EIT regime, since $\Omega_c<(\gamma_{20}-\gamma_{10})/2$, one can rewrite ${\rm Im}(\rho_{20})$ in Eq.~(\ref{eq:t}) as the sum of a broad {\it positive} Lorentzian and a narrow {\it negative} Lorentzian:
\begin{equation}
{\rm Im}(\rho_{20})_{\rm EIT}=\frac{C_+^2}{\delta^2+\gamma_+^2}-\frac{C_-^2}{\delta^2+\gamma_-^2},
\label{eq:eit}
\end{equation}
where $\gamma_{\pm}=\frac{1}{2}[\gamma_{20}+\gamma_{10}\pm\sqrt{(\gamma_{20}-\gamma_{10})^2-4\Omega_c^2}]$.
Note that Eq.~(\ref{eq:eit}) deviates from the sum of two {\it positive} Lorentzians, indicating that destructive interference occurs in this driven three-level system. As for raising the bound $(\gamma_{20}-\gamma_{10})/2$ by increasing $\gamma_{20}$ in the experiment, it is to have a broader range of $\Omega_c$ to obersve the EIT.
In the strong ATS regime with $\Omega_c\gg (\gamma_{20}-\gamma_{10})/2$, ${\rm Im}(\rho_{20})$ in Eq.~(\ref{eq:t}) is reduced to the sum of two {\it positive} equal-width but shifted Lorentzians:
\begin{equation}
{\rm Im}(\rho_{20})_{\rm ATS}=\frac{C^2}{(\delta-\delta_0)^2+\gamma^2}+\frac{C^2}{(\delta+\delta_0)^2+\gamma^2},
\label{eq:ats}
\end{equation}
where $\gamma=(\gamma_{20}+\gamma_{10})/2$, and $\delta_0=\frac{1}{2}\sqrt{4\Omega_c^2-(\gamma_{20}-\gamma_{10})^2}$.
In Figs.~\ref{figure3}(a)-3(d), where the corresponding driving strengths are the same as in Figs.~\ref{figure2}(b)-2(e), we directly compare our experimentally observed transmission spectrum with the EIT model in Eq.~(\ref{eq:eit}) and the ATS model in Eq.~(\ref{eq:ats}). For a weak control field, the observed transmission spectrum fits very well with the EIT model and deviates appreciably from the ATS model [see Figs.~\ref{figure3}(a) and 3(b)]. However, when the control field has a moderate strength, the transmission spectrum deviates appreciably from both the EIT and ATS models [see Fig.~\ref{figure3}(c)]. Furthermore, for a strong control field, the transmission spectrum fits very well with the ATS model and deviates drastically from the EIT model [see Fig.~\ref{figure3}(d)]. Note that when fitting with the EIT model in this strong control-field range, $T'$ can become negative. This further indicates that the EIT model cannot ever be applied. These comparisons clearly demonstrate that a transition from EIT to ATS occurs when increasing the strength of the control field applied to the three-level system.

We can qualitatively discriminate EIT from ATS by using Akaike's Information Criterion (AIC)~\cite{Peng2014,Anisimov2011}, which can identify the most informative model based on relative entropy. The information loss of a given model with $k$ fitting parameters to the experimental data is quantified by $I=N\ln(R/N)+2k$, where $N$ is the number of data points for fitting and $R$ denotes the fitting residual sum of squares. The per-point AIC contribution is given by $\bar{I}=I/N$.
In our experiment, each transmission spectrum contains $N=61$ points. By calculating AIC per-point weights $\bar{w}_{\rm EIT}$ and $\bar{w}_{\rm ATS}$:
\begin{equation}
\bar{w}_{\rm EIT}=\frac{\exp(-\frac{1}{2}\bar{I}_{\rm EIT})}{\exp(-\frac{1}{2}\bar{I}_{\rm EIT})+\exp(-\frac{1}{2}\bar{I}_{\rm ATS})},
\end{equation}
and $\bar{w}_{\rm ATS}=1-\bar{w}_{\rm EIT}$, one can determine whether the EIT model [Eq.~(\ref{eq:eit})] or ATS model [Eq.~(\ref{eq:ats})] is the most likely case for the experimental data. When the control-field driving strength is small enough ($\Omega_c<\Omega_{\rm EIT}=2\pi\times2.57$~MHz), the system is in the EIT regime and the EIT model can fit the experimental data extremely well, while the ATS model fits poorly [see Fig.~\ref{figure3}(e)]. When increasing the driving strength to $\Omega_c>\Omega_{\rm EIT}$, the system transitions from the EIT to ATS regime. Note that the ATS emerges when the driving strength $\Omega_c$ slightly exceeds $\Omega_{\rm EIT}$, but the system is still in the EIT-dominated regime. Thus, the EIT model fits with the experimental data better than ATS model does. While the driving strength $\Omega_c$ reaches around $\Omega_{\rm AIC}=2\pi\times4.28$~MHz, either model cannot fit well with the experimental data, so the system is in the {\it intermediate} or transition regime. For $\Omega_c>\Omega_{\rm AIC}$, when increasing $\Omega_c$, the ATS model fits with the experimental data increasingly better than the EIT model does. Therefore, the AIC per-point weights in Fig.~\ref{figure3}(e) further reveals that, for our tunable 3D transmon, EIT and ATS occurs in the weak and strong driving regimes, respectively, and a transition occurs between them.

\section {Discussion and conclusion}

Although observed and extensively studied in atomic systems, EIT as an important quantum-optics phenomenon has not been observed in macroscopic quantum systems such as superconducting quantum circuits. This is because the fulfillment of the conditions for realizing EIT in a superconducting circuit is much more difficult than in an atomic system. However, by tuning the transmon with an external magnetic field, we have successfully reached the EIT parameter regime of this superconducting circuit. From the experimental observations, we have also clearly identified the EIT and ATS regimes as well as the transition regime in between.

Theoretical studies give a threshold of the control-field drive strength at the border between EIT and ATS, which corresponds to $\Omega_{\rm EIT}=(\gamma_{20}-\gamma_{10})/2=2\pi\times2.57$~MHz in our experiment. In Fig.~\ref{figure3}(e), the EIT regime and the transition from EIT- to ATS-dominated regime are clearly shown. By fitting the calculated per-point weight with the experimental results, it is estimated that the noise in our experiment is about $3\%$ of the signal. The crossing point of the curves corresponds to the threshold $\Omega_{\rm AIC}=2\pi\times4.28$~MHz. When $\Omega_c < \Omega_{\rm AIC}$, the EIT model fits the experimental data better than the ATS model and vice versa when $\Omega_c > \Omega_{\rm AIC}$. Obviously, the threshold $\Omega_{\rm AIC}$ is larger than the threshold $\Omega_{\rm EIT}$ determined by the EIT theory. Indeed, when the driving strength $\Omega_c$ is slightly larger than $\Omega_{\rm EIT}$, the system starts to transition from the EIT to ATS regime, but it is still in the EIT-dominated regime (i.e., the EIT model describes the system better than the ATS model). Therefore, our experimental data demonstrates that the threshold $\Omega_{\rm AIC}$ describes the EIT-ATS transition better than the threshold $\Omega_{\rm EIT}$.

\section*{Acknowledgments}

This work is supported by the NSAF Grant Nos.~U1330201 and U1530401, the NSFC Grant Nos.~91421102 and 60836001, and the MOST 973 Program Grant Nos.~2014CB848700, 2014CB921401 and 2011CBA00304. F.N. is partly supported by the RIKEN iTHES Project, MURI Center for Dynamic Magneto-Optics, the ImPACT Program of JST, and Grant-in-Aid for Scientific Research (S). J.S.T. is partially supported by the Japanese Cabinet Office's ImPACT project. We thank Yu-xi Liu, P.-M. Billangeon for valuable discussions, K. Kusuyama for technical assistance, and Anton F. Kockum for a critical reading of the manuscript.

{\it Note added}. Recently, we became aware of a work by Novikov {\it et al.}~\cite{Novikov}, which also studies EIT in a similar setup.

\begin{figure}
\begin{center}
\includegraphics[scale=0.9]{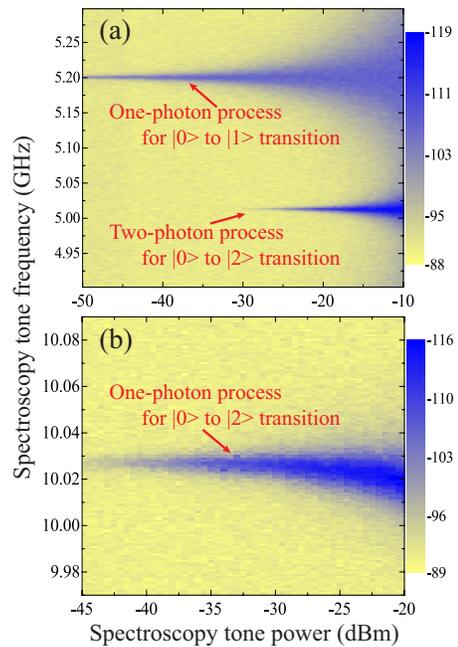}
\caption{(color online) The transition spectroscopy between states $|0\rangle$ and $|2\rangle$ of the tunable 3D transmon at a flux bias giving $E_J/E_C=27.5$. (a) One-photon transition process between states $|0\rangle$ and $|1\rangle$, and two-photon transition process between states $|0\rangle$ and $|2\rangle$. (b) One-photon transition process between states $|0\rangle$ and $|2\rangle$.}\label{fig4}
\end{center}
\end{figure}

\section*{Appendix: One-photon transition between states $|0\rangle$ and $|2\rangle$}

In the experiment, we have also measured the one-photon transition spectroscopy between the eigenstates $|0\rangle$ and $|2\rangle$ of the 3D transmon at another flux bias giving $E_J/E_C=27.5$. Figure~\ref{fig4}(a) shows the transition spectroscopy for the one-photon process between the eigenstates $|0\rangle$ and $|1\rangle$ of the transmon as well as the two-photon process between the eigenstates $|0\rangle$ and $|2\rangle$. The corresponding transition frequencies are measured to be $\omega_{10}/2\pi=5.202$ GHz and $\frac{1}{2}\omega_{20}/2\pi=5.014$ GHz, respectively. For comparison, the transition spectroscopy for the one-photon process between states $|0\rangle$ and $|2\rangle$ is also shown in Fig.~\ref{fig4}(b), with the resonance exactly at the transition frequency $\omega_{20}/2\pi=10.028$~GHz. Note that these transition frequencies are still far detuned from the cavity frequency ($8.2169$ GHz). Moreover, as shown in Fig.~\ref{fig5}, we have measured the Rabi oscillations between states $|0\rangle$ and $|2\rangle$ by using a driving field with frequency $10.028$~GHz. It gives that the Rabi oscillation period is $56.8$~ns and the oscillation decay time is $130.6\mathrm{ns}$ by fitting with the experiment. These results reveal that one-photon transition between $|0\rangle$ and $|2\rangle$ indeed occurs in our 3D transmon system.

\begin{figure}
\includegraphics[scale=0.5]{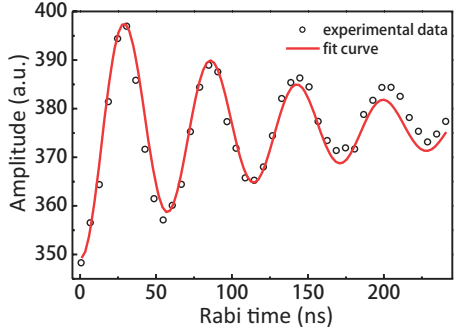}\label{fig5}
\caption{(color online) The measured Rabi oscillations between states $|0\rangle$ and $|2\rangle$ of the 3D trnasmon by using a $10.028 \mathrm{GHz}$ driving field. The black circles are experimental data and the red curve is an exponentially damped sinusoidal fit. The Rabi oscillation period is $56.8$~ns and the oscillation decay time is $130.6$~ns from the fit.}\label{fig5}
\end{figure}

\begin{figure}
\begin{center}
\includegraphics[scale=1.1]{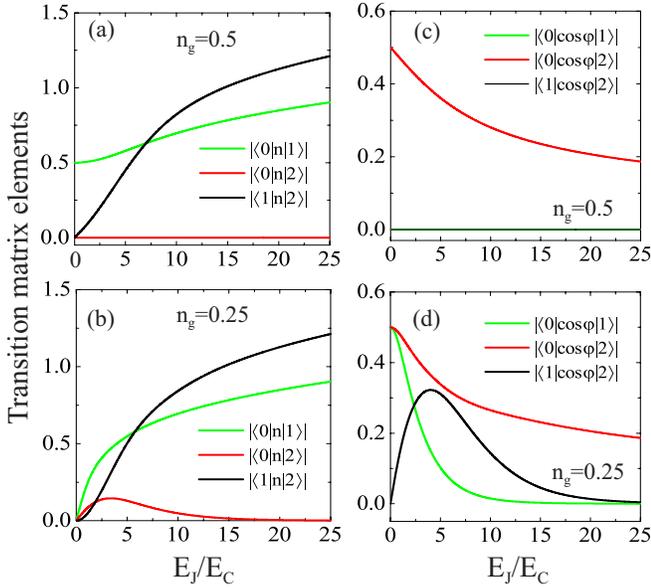}
\caption{(color online) Electric- and magnetic-dipole transition matrix elements of the 3D transmon as a function of the ratio $E_J/E_C$. (a) and (b) correspond to the electric-dipole one-photon transitions for $n_g=0.5$ and $0.25$, respectively. (c) and (d) correspond to the magnetic-dipole one-photon transitions for $n_g=0.5$ and $0.25$, respectively.  }\label{fig6}
\end{center}
\end{figure}

The Hamiltonian of the transmon can be written as
\begin{equation}
H=4E_c(n-n_g)^2-E_J\cos\varphi.
\label{tranH}
\end{equation}
The electric-dipole transition matrix element between eigenstates $|i\rangle$ and $|j\rangle$ of the transmon is proportional to $\langle i|n|j\rangle$, due to the field-induced time-dependent variation of the charge bias $n_g$. For the 3D transmon, the charge bias $n_g$ is not fixed but floated, so it can change for a different experimental setup. In Fig.~\ref{fig6}(a), we calculate $|\langle 0|n|1\rangle|$, $|\langle 0|n|2\rangle|$ and $|\langle 1|n|2\rangle|$ at $n_g=0.5$. These calculated transition matrix elements are similar to the results obtained in Ref.~\cite{Koch}. The result of $|\langle 0|n|2\rangle|=0$ indicates that no electric-dipole one-photon transition occurs between states $|0\rangle$ and $|2\rangle$. When $n_g\neq 0.5$ (e.g., $n_g=0.25$), $|\langle 0|n|2\rangle|$ can be nonzero, but it is much smaller than $|\langle 0|n|1\rangle|$, especially around $E_J/E_c=16.99$ (i.e., the ratio given in our experiment for the EIT) [see Fig.~\ref{fig6}(b)]. Therefore, the electric-dipole one-photon transition between $|0\rangle$ and $|2\rangle$ is still too weak in the case of $n_g\neq 0.5$.

For the 3D transmon used in our experiment, the single Josephson junction is replaced by a symmetric SQUID with an effective Josephson coupling
\begin{equation}
E_J=2E_{J0}\cos\left(\frac{\pi\Phi_x}{\Phi_0}\right),
\end{equation}
where $E_{J0}$ is the Josephson coupling energy of each junction in the SQUID.
In addition to the static magnetic flux $\Phi_x$, when a weak time-dependent magnetic flux $\Phi_a(t)$ is applied (i.e., $|\pi\Phi_a(t)/\Phi_0|\ll 1$), the Hamiltonian becomes
\begin{equation}
H(t)=H+I\Phi_a(t),
\end{equation}
where
\begin{equation}
I=\frac{2\pi E_{J0}}{\Phi_0}\sin\left(\frac{\pi\Phi_x}{\Phi_0}\right)\cos\varphi
\end{equation}
is the circulating current in the SQUID loop~\cite{You2001}. The magnetic-dipole transition matrix element between eigenstates $|i\rangle$ and $|j\rangle$ of the transmon is proportional to $\langle i|\cos\varphi|j\rangle$. In Figs.~\ref{fig6}(c) and (d), we show the calculated $|\langle 0|\cos\varphi|1\rangle|$, $|\langle 0|\cos\varphi|2\rangle|$ and $|\langle 1|\cos\varphi|2\rangle|$. At $n_g=0.5$, while both $|\langle 0|\cos\varphi|1\rangle|=0$ and $|\langle 1|\cos\varphi|2\rangle|=0$, $|\langle 0|\cos\varphi|2\rangle|$ is nonzero [see Fig.~\ref{fig6}(c)], indicating that only the magnetic-dipole one-photon transition between eigenstates $|0\rangle$ and $|2\rangle$ is allowed in this case. When $n_g\neq 0.5$ (e.g., $n_g=0.25$), both $|\langle 0|\cos\varphi|1\rangle|$ and $|\langle 1|\cos\varphi|2\rangle|$ become nonzero, but they are smaller than $|\langle 0|\cos\varphi|2\rangle|$. In particular, they are much smaller than $|\langle 0|\cos\varphi|2\rangle|$ at $E_J/E_c=16.99$ [see Fig.~\ref{fig6}(d)].
Because $E_J\gg E_c$ in the transmon, when the field-induced time-dependent variation of the charge bias $n_g$ is small, the magnetic-dipole one-photon transition between $|0\rangle$ and $|2\rangle$ can become as important as the electric-dipole one-photon transitions $|0\rangle\rightarrow|1\rangle$ and $|1\rangle\rightarrow|2\rangle$. This explains the observation of the one-photon transition between states $|0\rangle$ and $|2\rangle$ in our tunable 3D transmon.


\end{document}